\documentclass[a4paper,11pt]{article}
\usepackage{pos}
\renewcommand{\logo}{\relax}
\usepackage{booktabs}

\usepackage{array}
\usepackage{tikz}
\usetikzlibrary{arrows.meta, decorations.markings,decorations.pathmorphing,intersections,patterns}
\tikzset{->-/.style={
        postaction={decorate},
        decoration={markings, mark=at position .525 with {\arrow{>}}}
    },
  ->>-/.style={
    postaction={decorate},
    decoration={
      markings,
      mark=at position .475 with {\arrow{>}},
      mark=at position .575 with {\arrow{>}}
    }
  }}

\newcommand{\pinm}{\mathrm{Pin}^{-}}
\newcommand{\pinp}{\mathrm{Pin}^{+}}

\newcommand{\ZZ}{\mathbb{Z}}
\newcommand{\RPt}{\mathbb{R}\mathrm{P}^2}

\DeclareMathOperator{\sign}{\mathrm{sign}}
\DeclareMathOperator{\Pf}{\mathrm{Pf}}

\title{How to formulate the $\mathbb{Z}_8$
 topological invariant of Majorana fermion on the lattice}

\author*{Sho Araki}
\author{Hidenori Fukaya}
\author{Tetsuya Onogi}
\author{Satoshi Yamaguchi}

\affiliation{Department of physics, The University of Osaka \\ Toyonaka 560-0043, Japan}

\emailAdd{araki@het.phys.sci.osaka-u.ac.jp}
\emailAdd{hfukaya@het.phys.sci.osaka-u.ac.jp}
\emailAdd{onogi@het.phys.sci.osaka-u.ac.jp}
\emailAdd{yamaguchi@het.phys.sci.osaka-u.ac.jp}

\abstract{Topological invariants and their associated anomalies have played a crucial role in understanding low-energy phenomena in quantum field theories. In lattice gauge theory, the standard $\mathbb{Z}$-valued Atiyah-Singer index is formulated via the overlap Dirac operator through the Ginsparg-Wilson relation, but extensions to more general topological invariants have remained limited. In this work, we propose a lattice formulation of the Arf-Brown-Kervaire (ABK) invariant, which takes values in $\mathbb{Z}_8$. The ABK invariant arises in Majorana fermion partition functions with reflection symmetry on two-dimensional non-oriented manifolds, and its definition involves an infinite sum over Dirac eigenvalues that must be properly regularized. By carefully treating the boundary conditions, with and without a domain-wall mass term, we demonstrate that the ABK invariant can be extracted from Pfaffians of the Wilson Dirac operator. We further provide numerical verification on two-dimensional lattices, showing that the 
$\mathbb{Z}_8$-valued results on the torus, Klein bottle, real projective plane, and Möbius strip agree with those in the continuum theory.

OU-HET-1303}

\FullConference{The 42nd International Symposium on Lattice Field Theory (LATTICE2025)\\
2-8 November 2025\\
Tata Institute of Fundamental Research, Mumbai, India\\}

\begin{document}
\maketitle

\section{Introduction}
Topology plays important roles in understanding nonperturbative aspects of quantum field theories such as instantons, anomalies and SPT phases \cite{Witten:2015aba}. In lattice field theory, which is a powerful tool for nonperturbative analyses of QFT, topology is difficult property because of absence of continuity, although we expected to recovered in the continuum limit.

However, the index of Dirac operators, which is equal to the topological number of the gauge field in continuum theory \cite{Atiyah:1968}, can be formulate on the lattice.
It was shown in \cite{Neuberger:1997fp,Hasenfratz:1998ri} that the index of lattice Dirac operators
satisfying the Ginsparg-Wilson relation \cite{Ginsparg:1981bj}
can be formulated in a similar way to the continuum theory;
the index is defined by the Dirac operator zero modes, thanks to the exact chiral symmetry on a lattice \cite{Luscher:1998pqa}.

In resent studies, the Ginsparg Wilson relation was generalized \cite{Clancy:2023ino} to wider fermionic systems, including those related to discrete symmetries or real structures (e.g. mod-2 type indices), was formulated on lattice.
In \cite{Aoki:2024sjc,Aoki:2025gca}, another new generalization of the indices was proposed. By employing $K$-theory, it is shown that the spectral flow of the massive Wilson Dirac operators coincide with the index of Overlap Dirac operators. 
Since the massive Dirac operators are used, any chiral symmetry or Ginsparg-Wilson relation are not required.
This formulation of index can be extended to not only usual index or mod-2 type index but also the Atiyah-Patodi-Singer(APS) index \cite{APS}, originally defined on manifolds with boundary, by using domain-wall fermion approach.

There still exist topological invariants not captured by these methods. In some fermionic system, topological invariants valued in $\ZZ_8$ or $\ZZ_{16}$ arise \cite{Kapustin:2014dxa} which are not expressed as indices of Dirac operators.
Furthermore, fully capturing such topological invariants requires to consider fermions on a variety of manifold, including non-orientable cases\footnote{On these manifolds, the overlap operator and its index is not formulated well.
In particular, the Ginsparg-Wilson relation is known to be broken
when we impose nontrivial boundary conditions \cite{Luscher:1998pqa}.
}. 

In this work, we try to formulate a $\ZZ_8$-valued topological invariant, known as the Arf-Brown-Kervaire (ABK) invariant \cite{Brown,Debray:2018wfz} on lattice fermion systems. The ABK invariant appears in the complex phase of the partition function of the two-dimensional Majorana fermion with the reflection symmetry\cite{Kaidi:2019tyf}.
Mainly, we examine that the ABK invariant can be reproduced by explicitly computing the partition function of Wilson fermion.
Furthermore, since detecting the ABK invariant essentially requires considering non-orientable manifolds, we also provide a method to realize such situations on the lattice. 
The full details of our work is in \cite{Araki:2025xly}.

\section{2D Majorana fermion and the ABK invariant}
To see how the ABK invariant appears in the theory, we first consider the continuum case. 
We work in Euclidean spacetime and take gamma matrices to satisfy $\left\{ \gamma^a, \gamma^b \right\} = \delta^{ab}$.
Majorana fermions are obtained by imposing on the standard Dirac fermion $\psi$,
the Majorana condition $\bar{\psi}=\psi^T C$ with the charge conjugation matrix $C$ satisfying $C\gamma^a C^{-1} = -(\gamma^a)^T$.
The action of a Majorana fermion on a two-dimensional manifold $X$ is given by
\begin{equation}
  S =   \frac{1}{2}\int_{X}^{} d^2x \psi^T C (D + m) \psi, \label{eq:continuum_action}
\end{equation}
where $D=\gamma^\mu D_\mu$ is the Dirac operator.
Here we impose that the Lagrangian density is invariant under reflection,
\begin{equation}
  \psi(x,y) \to R_x \psi(x,y) = \gamma^1 \gamma^3 \psi(-x,y) \label{eq:reflection}. 
\end{equation}
Including this reflection transformation in the spinor rotation group $\mathrm{Spin}(2)$ yields $\pinm(2)$, where the "$-$" originates from $R_x^2 = -1$.
Imposing this requirement allows the action to be well-defined on a non-orientable manifold since the Lagrangian density is the same in orientation-reversed coordinate systems.
More precisely, Majorana fermions can be formulated on manifolds which is equipped with a $\pinm$ structure\footnote{This structure is required for a globally well-defined spinor field. It is not unique; different choices exist, distinguished by sign of boundary condition when fermions going once around the manifold.}.

In that setup, the partition function of the Majorana fermion may have a nontrivial complex phase\footnote{
   We use a regularization (like the Pauli-Villars regularization), such that no phase appears when $m>0$ .
}. Taking $m\to -\infty$, it exhibits a quantized phase\cite{Witten:2015aba}\cite{Kaidi:2019tyf}
\begin{equation}
  Z(X,s;m) \propto \exp\left(\frac{i\pi}{4}\beta\right). \label{eq:Z8phase}
\end{equation}
Here, $\beta$ takes valued in $0,1,\cdots,7$ and depends only on the topological backgrounds: the topology of the manifold $X$ and its $\pinm$ structure $s$. This $\ZZ_8$ topological invariant is called the Arf-Brown-Kervaire invariant\cite{Brown}.

The appearance of the $\ZZ_8$ reflects an eightfold classification of fermionic SPT phases with the reflection symmetry \cite{Fidkowski:2009dba,Fidkowski:2010jmn}. In other words, there are eight distinct vacua that cannot deform continuously to each other preserving the symmetry and the mass gap.

This analysis can be extended to manifolds with boundary as well.
To realize open manifolds, we adopt a domain-wall mass term \cite{Kaplan:1992bt} rather than imposing boundary conditions.
The domain-wall mass term to realize an open manifold $X$ is given by  
\begin{equation}
  m_X(x,y) = \begin{cases}
    -m_0 & \text{if } (x,y) \text{ in } X, \\
    +m_0 & \text{otherwise}
  \end{cases},
\end{equation}
where we assume that $m_0>0$. At the large mass limit $m_0 \to \infty$, the phase of partition function is quantized in $\ZZ_8$ as is in closed manifolds case. This provides a formulation of the ABK invariant for open manifolds. The phase quantization for open manifolds is connected to the anomaly inflow mechanism of the edge fermion that effectively appears on the domain wall. See \cite{Witten:2015aba} for more details.

\section{The ABK invariant on the lattice}
The purpose of this work is to formulate the lattice ABK invariant by using the path integral of massive fermion\footnote{
In previous studies, a state sum combinatorial definition of the ABK
invariant 
was formulated given by \cite{Kobayashi:2019xxg,Inamura:2019hpu}. We also refer the readers to
\cite{Shapourian:2016kvr,Shiozaki:2017ive} in which the ABK invariant
was discussed using the Hamiltonian formalism. 
}.
Unlike the Dirac operator index, the ABK invariant cannot be obtained from a definite number of eigenvalues. Instead, we compute the partition function of Majorana fermions directly and extract the lattice ABK invariant from its phase. At finite volume, the partition function is the Pfaffian of a finite-size matrix, so it becomes possible to be obtained by numerical calculation.

Imposing twisted (orientation-reversing) boundary conditions to fermions, we obtain various types of manifolds including non-orientable ones such as the Klein Bottle or the real projective plane. We employ the Wilson fermions to realize these manifolds.
We emphasize that this approach does not rely on imposing chiral symmetry.
\subsection{Lattice setup}
We consider a standard two-dimensional square lattice whose size is denoted by $N_x a\times N_y a$
with the lattice spacing $a$.
The action is given by
\begin{equation}
  S_{\text{W}} = \sum_{x,y} a^2 \frac{1}{2} \psi(x,y)^T C D_W(m) \psi(x,y),
\end{equation}
where $x$ and $y$ take integer multiple of $a$: $0,a,2a\cdots (N_{x/y}-1)a$, 
and $D_W(m)$ denotes the massive Wilson-Dirac operator
\begin{equation}
  D_W(m)=
\sum_\mu \gamma^\mu \frac{\nabla^f_\mu + \nabla^b_\mu}{2}+ m + \sum_\mu \frac{\nabla^f_\mu - \nabla^b_\mu}{2},
\end{equation}
where $\nabla^{f}_\mu$ and $\nabla^{b}_\mu$ are the forward and backward difference operators, respectively.

The partition function of this lattice system can be written by the Pfaffian, as 
\begin{equation}
  Z(X,s;m) = \mathrm{Pf}(C D_W(m)).
\end{equation}
We define the lattice version of the ABK invariant $\beta^\text{latta}$ by the
phase of the partition function
\begin{equation}
  \frac{\mathrm{Pf}(C D_W(m))}{\mathrm{Pf}(C D_W(|m|) )} \propto \exp\left(i\frac{2\pi}{8} \beta^{\text{latt}}\right).
\end{equation}
Here, the Pauli-Villar's partition function $\mathrm{Pf}\left( C D_{W}(|m|)\right)=Z(|m|)$ is introduced 
to compare the results with the continuum theory. 
In this formulation, $\beta^\text{latt}$ is not guaranteed to be integer-valued. 

Whether the correct integer is recovered in the continuum limit must
be verified by explicit
computations on a variety of manifolds, where we take the continuum limit ($a \to 0)$ as well as the large fermion mass limit ($m \to -\infty$).
In this work we fix the physical lattice size $L_x =N_x a, L_y= N_y a$ and the
theory is controlled by the dimensionless parameters $ma$ and $N_{x/y}$.
In order to take these limit, we typically set $ma=-1,$ and take the $ N_{x/y} \to \infty$ limit
in the following calculation.

\subsection{Construction of manifolds}
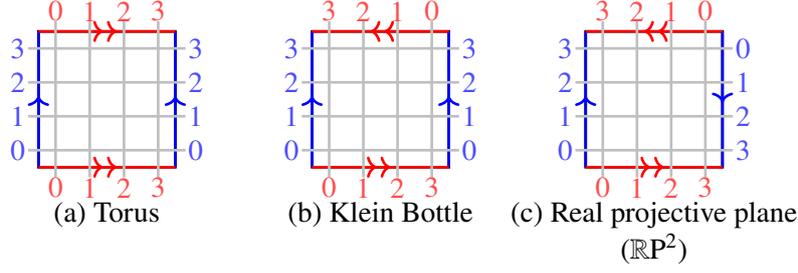
\begin{figure}[htbp]
    \centering
    \begin{tikzpicture}[scale=0.9, line width=1pt
]

    \node at (1,-0.7) {(a) Torus};
    \draw (0,0) rectangle (2,2);

    \draw[->-,blue] (0,0) -- (0,2);
    \draw[->-,blue] (2,0) -- (2,2);

    \draw[->>-,red] (0,0) -- (2,0);
    \draw[->>-,red] (0,2) -- (2,2);
    \def\N{4}          
    \def\a{0.5}        
    \def\LDx{0.25}
    \def\LDy{0.25}
    \def\margin{0.4}  
    \def\latticeColor{gray!50}
        
    \foreach \j in {0,...,\numexpr\N-1\relax}{
      \draw[\latticeColor] (-\margin+\LDx,\j*\a+\LDy)--(\N*\a-1*\a+\margin+\LDx,\j*\a+\LDy);
    }
        
    \foreach \i in {0,...,\numexpr\N-1\relax}{
      \draw[\latticeColor] (\i*\a+\LDx, -\margin+\LDy) -- (\i*\a+\LDx, \a*\N-\a + \margin+\LDy);
    }
    \node[blue!70] at (-0.3,0.25) {0};
    \node[blue!70] at (-0.3,0.75) {1};
    \node[blue!70] at (-0.3,1.25) {2};
    \node[blue!70] at (-0.3,1.75) {3};
    \node[blue!70] at (2.3,0.25) {0};
    \node[blue!70] at (2.3,0.75) {1};
    \node[blue!70] at (2.3,1.25) {2};
    \node[blue!70] at (2.3,1.75) {3};
    \node[red!70] at (0.25,-0.3) {0};
    \node[red!70] at (0.75,-0.3) {1};
    \node[red!70] at (1.25,-0.3) {2};
    \node[red!70] at (1.75,-0.3) {3};
    \node[red!70] at (0.25,2.3) {0};
    \node[red!70] at (0.75,2.3) {1};
    \node[red!70] at (1.25,2.3) {2};
    \node[red!70] at (1.75,2.3) {3};
    \begin{scope}[xshift=4.0cm]
        \node at (1,-0.7) {(b) Klein Bottle};
        \draw (0,0) rectangle (2,2);

        \draw[->-,blue] (0,0) -- (0,2);
        \draw[->-,blue] (2,0) -- (2,2);

        \draw[->>-,red] (0,0) -- (2,0);
        \draw[->>-,red] (2,2) -- (0,2);    
        \def\N{4}          
        \def\a{0.5}        
        \def\LDx{0.25}
        \def\LDy{0.25}
        \def\margin{0.4}  
        \def\latticeColor{gray!50}
            
        \foreach \j in {0,...,\numexpr\N-1\relax}{
          \draw[\latticeColor] (-\margin+\LDx,\j*\a+\LDy)--(\N*\a-1*\a+\margin+\LDx,\j*\a+\LDy);
        }
            
        \foreach \i in {0,...,\numexpr\N-1\relax}{
          \draw[\latticeColor] (\i*\a+\LDx, -\margin+\LDy) -- (\i*\a+\LDx, \a*\N-\a + \margin+\LDy);
        }
        \node[blue!70] at (-0.3,0.25) {0};
        \node[blue!70] at (-0.3,0.75) {1};
        \node[blue!70] at (-0.3,1.25) {2};
        \node[blue!70] at (-0.3,1.75) {3};
        \node[blue!70] at (2.3,0.25) {0};
        \node[blue!70] at (2.3,0.75) {1};
        \node[blue!70] at (2.3,1.25) {2};
        \node[blue!70] at (2.3,1.75) {3};
        \node[red!70] at (0.25,-0.3) {0};
        \node[red!70] at (0.75,-0.3) {1};
        \node[red!70] at (1.25,-0.3) {2};
        \node[red!70] at (1.75,-0.3) {3};
        \node[red!70] at (0.25,2.3) {3};
        \node[red!70] at (0.75,2.3) {2};
        \node[red!70] at (1.25,2.3) {1};
        \node[red!70] at (1.75,2.3) {0};
    \end{scope}

    \begin{scope}[xshift=8.0cm]
        \node at (1,-0.7) {(c) Real projective plane};
        \node at (1,-1.2) {($\RPt$)};
        \draw (0,0) rectangle (2,2);
        
        \draw[->-,blue] (0,0) -- (0,2);
        \draw[->-,blue] (2,2) -- (2,0);

        \draw[->>-,red] (0,0) -- (2,0);
        \draw[->>-,red] (2,2) -- (0,2);    
        \def\N{4}          
        \def\a{0.5}        
        \def\LDx{0.25}
        \def\LDy{0.25}
        \def\margin{0.4}  
        \def\latticeColor{gray!50}
            
        \foreach \j in {0,...,\numexpr\N-1\relax}{
          \draw[\latticeColor] (-\margin+\LDx,\j*\a+\LDy)--(\N*\a-1*\a+\margin+\LDx,\j*\a+\LDy);
        }
            
        \foreach \i in {0,...,\numexpr\N-1\relax}{
          \draw[\latticeColor] (\i*\a+\LDx, -\margin+\LDy) -- (\i*\a+\LDx, \a*\N-\a + \margin+\LDy);
        }
        \node[blue!70] at (-0.3,0.25) {0};
        \node[blue!70] at (-0.3,0.75) {1};
        \node[blue!70] at (-0.3,1.25) {2};
        \node[blue!70] at (-0.3,1.75) {3};
        \node[blue!70] at (2.3,0.25) {3};
        \node[blue!70] at (2.3,0.75) {2};
        \node[blue!70] at (2.3,1.25) {1};
        \node[blue!70] at (2.3,1.75) {0};
        \node[red!70] at (0.25,-0.3) {0};
        \node[red!70] at (0.75,-0.3) {1};
        \node[red!70] at (1.25,-0.3) {2};
        \node[red!70] at (1.75,-0.3) {3};
        \node[red!70] at (0.25,2.3) {3};
        \node[red!70] at (0.75,2.3) {2};
        \node[red!70] at (1.25,2.3) {1};
        \node[red!70] at (1.75,2.3) {0};
    \end{scope}

    \end{tikzpicture}
    \caption{Identification patterns of the link variables for a torus, Klein bottle, and real projective plane are shown. The links across the boundaries with the same labels are identified.}
    \label{fig:manifolds_identification}
\end{figure}

To construct lattice fermions on various manifolds, we begin with an $N_x a\times N_y a$ square lattice and then identify (glue) its edges rather than using a triangular-lattice approach \cite{Brower:2016vsl,Burda:1999aj}. 
By introducing the operation of gluing opposite edges after reversing their orientations, it becomes possible to realize lattice networks for non-orientable manifolds such as the Klein bottle and two-dimensional real projective plane ($\RPt$), as Fig.~\ref{fig:manifolds_identification}.
For the fermion field, we impose boundary conditions corresponding to the gluing.
In particular, for an orientation-reversing identification, we glue fermions field after the reflection operations $R_x,R_y$ in \eqref{eq:reflection}.
Here we note that there is a sign choice in the fermion boundary conditions. This corresponds to a choice of the $\pinm$ structure.

The flat torus (Fig.~\ref{fig:manifolds_identification}(a)) is obtained by identifying opposite edges without reversing their orientation.
For each of two directions ($x$ and $y$), there are two possible boundary conditions of fermions:
periodic ($P$) or antiperiodic ($A$).
Specifically they are given by
\begin{align}
\psi\left(N_xa,y\right) = \pm \psi\left(0,y\right), 
 && \psi\left(x,N_y a\right) = \pm \psi\left(x,0\right).\label{eq:TorusBC}
\end{align}
There are four different $\pinm$ structures on the torus, $PP$, $PA$, $AP$ and $AA$.

The Klein bottle is obtained by gluing the upper and lower edges of 
a cylinder after reversing the orientation (Fig.~\ref{fig:manifolds_identification}(b)). 
First we identify $x=0$ and $x=N_x a$, 
with $P$ or $A$ boundary conditions:
\begin{equation}
  \psi\left(N_x a,y\right) = \pm \psi\left(0,y\right) \label{eq:KBBC_x}.
\end{equation}
Then we glue the $y = 0$ with $y = N_y a$ edges by orientation reflecting boundary condition\footnote{A similar approach for considering fermion partition function on non-orientable manifolds is represented in \cite{Mages:2015scv}.},
\begin{equation}
\psi\left(x,N_y a\right) = s R_x\psi(x,0)= s \gamma^1 \bar{\gamma}\psi\left((N_x-1)a-x,0\right),
\label{eq:KBBC_y}
\end{equation}
where we have two choices of the sign $s=\pm 1$. 
Hence, the Klein bottle admits four distinct $\pinm$ structures we denote by 
$P+,P-,A+$ and $A-$, where the $\pm$ corresponds to the sign $s$ of Eq.~\eqref{eq:KBBC_y}.

In order to consider Majorana fermions on an $\RPt$, we  glue both the top and bottom 
and the left and right boundaries of the lattice with a twist in each direction (see Fig.~\ref{fig:manifolds_identification} (c)).
Explicit forms of the boundary conditions are 
\begin{align}
     \psi\left(N_x a,y\right)= s_x R_y\psi(0,y) &&
   \psi\left(x,N_y a\right) = s_y R_x\psi(x,0).
\end{align}
Naively, the choices $s_x =\pm$ and $s_y =\pm$ suggest four $\pinm$ structures. However, only two of these,$(s_x,s_y)=(+,-)$ or $(-,+)$, are consistent with the continuum $\RPt$. This can be checked by focusing on the fermion rotation at a corner of the lattice (for the correct $\pinm $ structures, the fermion acquires $-1$ factor under a $2\pi$ rotation.), or from a discussion of the curvature at that point.

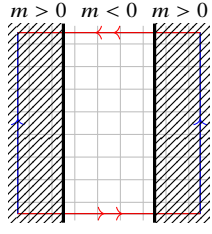
\begin{figure}[htbp]
    \centering
    \begin{tikzpicture}[scale=0.6]
   
        \def\N{8}          
        \def\a{0.5}        
        \def\LDx{0.25}
        \def\LDy{0.25}
        \def\margin{0.4}  
        \def\latticeColor{gray!50}
            
        \foreach \j in {0,...,\numexpr\N-1\relax}{
          \draw[\latticeColor] (-\margin+\LDx,\j*\a+\LDy)--(\N*\a-1*\a+\margin+\LDx,\j*\a+\LDy);
        }
            
        \foreach \i in {0,...,\numexpr\N-1\relax}{
          \draw[\latticeColor] (\i*\a+\LDx, -\margin+\LDy) -- (\i*\a+\LDx, \a*\N-\a + \margin+\LDy);
        }

        \draw (0,0) rectangle (4,4);

        \draw[->-,blue] (0,0) -- (0,4);
        \draw[->-,blue] (4,0) -- (4,4);

        \draw[->>-,red] (0,0) -- (4,0);
        \draw[->>-,red] (4,4) -- (0,4); 
        \fill[pattern=north east lines, pattern color=black] (-0.2,-0.2) rectangle (1,4.2);
        \fill[pattern=north east lines, pattern color=black] (3,-0.2) rectangle (4.2,4.2);
        \draw[very thick] (1,-0.2) -- (1,4.2);
        \draw[very thick] (3,-0.2) -- (3,4.2);
        \node[fill=white, inner sep=2pt,font=\scriptsize] at (0.45,4.5) {$m>0$};
        \node[fill=white, inner sep=2pt,font=\scriptsize] at (2,4.5) {$m<0$};
        \node[fill=white, inner sep=2pt,font=\scriptsize] at (3.55,4.5) {$m>0$};
    \end{tikzpicture}
    \caption{ Domain-wall mass configuration to realize a Möbius strip (the white region).}
    \label{fig:mobiusDW}
\end{figure}

To obtain  open manifolds with boundary, we employ the domain-wall
fermion construction.
The Möbius strip can be constructed by the following domain-wall mass term on the Klein Bottle:
\begin{equation}
  m_{\mathrm{MS}}(x,y) = \begin{cases}
    -m_0 & \text{if } \frac{1}{4}N_y a \leq y < \frac{3}{4}N_y a, \\
    +m_0 & \text{otherwise}
  \end{cases}. \label{eq:DWmassMS}
\end{equation}
As shown in Fig.\ref{fig:mobiusDW}, the 
$m<0$ region\footnote{Since the outer region carries the same $+$ sign as the Pauli-Villars field, it can be regarded as the exterior.} has the topology of a Möbius strip. 
The Möbius strip admits two $\pinm$ structures. They are distinguished by the sign choice of the $y$-direction boundary condition in the corresponding Klein bottle construction in Eq.\eqref{eq:KBBC_y}.
\subsection{Analytic computation}
On manifolds with translational invariance, the momentum-space description of the Dirac operator and its eigenvalues enable us the Fourier analysis of the
Majorana fermion Pfaffian. Here, we demonstrate a computation of the ABK invariant using the eigenvalues for such manifolds (the Torus and Klein bottle).

In the momentum space, the Wilson-Dirac operator $D_W(m)$ is expressed as
\begin{equation}
   i\gamma^1 \frac{1}{a}\sin(a p_x)+i\gamma^2 \frac{1}{a}\sin(a p_y)+ m + \frac{1}{a} \left( 2 -\cos(a p_x)-\cos(ap_y) \right) ,
\end{equation}
and it has eigenvalues
\begin{equation}
  \label{eq:eigenvalues}
  \mu^{\pm}(p_x,p_y)=\pm i \frac{1}{a}\sqrt{\sin(a p_x)^2+\sin(a p_y)^2} + m + \frac{1}{a} \left( 2-\cos(a p_x)-\cos(a p_y) \right).
\end{equation}
The allowed values of $p_x,p_y$ and $\sigma=\pm$ are determined by the boundary conditions and the lattice size. Once the entire spectrum is known, the partition function can be written as
\begin{equation}
  Z = \Pf(CD_W(m))={\prod_{p_x,p_y,\pm}^{}}'\mu^{\pm}(p_x,p_y),
\end{equation}
where ${\prod}'$ takes one eigenvalue from the doubly degenerate eigenvalues (this degeneracy follows from the Majorana property).

\subsubsection{Torus}
Under the periodic/anti-periodic boundary conditions~\eqref{eq:TorusBC}, momenta take values in integer/half-integer times the unit momenta,
\begin{equation}
  p_{x/y} = \begin{cases}
  \frac{2\pi}{aN_{x/y}}k & (k=0,\dots,N_{x/y}) \quad \text{  if periodic,} \\
  \frac{2\pi}{aN_{x/y}}\left(k+\frac{1}{2}\right) & (k=0,\dots,N_{x/y}) \quad \text{  if antiperiodic}
  \end{cases}.
\end{equation}
And both the eigenvalues $\mu^+(p_x,p_y),\mu^-(p_x,p_y)$ appear in the spectrum.

For most $(p_x,p_y)$, $\mu^+(p_x,p_y)$ and $\mu^-(p_x,p_y)$
are distinct eigenvalues, and their contributions to the phase of the partition function cancel each other.
However, for $p_x=p_y=0$, only one of doubly degenerate eigenvalues $\mu^+(0,0)=\mu^-(0,0)=m$ contribute to the phase, yielding a nontrivial factor $Z(T^2,PP,m) \propto \sign (m)$.
Therefore, the phase of partition function $Z(T^2,s;m)$ is nontrivial only when $s=PP$ and $m<0$.
By taking $m<0$, we obtain
\begin{align}
  \beta^{\mathrm{latt}}(T^2,PP;m) &= 4, \\
  \beta^{\mathrm{latt}}(T^2,PA;m) &=\beta^{\mathrm{latt}}(T^2,AP;m)=\beta^{\mathrm{latt}}(T^2,AA;m)= 0,
\end{align}
as the lattice ABK invariant.
This result is consistent with the ABK invariant in continuum theory.

\subsubsection{The Klein bottle}
When the orientation-reversing boundary condition Eq.~\eqref{eq:KBBC_y} is imposed, the allowed pairs of $(p_x,p_y,\sigma)$ are restricted.
The momenta takes values in
\begin{align}
  p_{x} &= \begin{cases}
  \frac{2\pi}{N_{x} a}k_x & (k_x=0,\dots,\lfloor \frac{N_x}{2} \rfloor) \quad \text{  if periodic} \\
  \frac{2\pi}{N_{x} a}\left(k_x+\frac{1}{2}\right) & (k_x=0,\dots,\lfloor \frac{N_x-1}{2} \rfloor) \quad \text{  if antiperiodic },
  \end{cases} \label{eq:KBpxs}\\
  p_y &= \frac{\pi}{N_y a}\left(k_y+\frac{1}{2}\right) \quad (k_y=0,\dots,2N_y-1).  \label{eq:KBpys}
\end{align}
Here, for $p_x \ne 0$ \footnote{Strictly speaking, $p_x=\pi/a$ case (which corresponds to a doubler mode) also should be treated as $p_x=0$ case. However it can be shown that this case does not contribute to the entire phase. }, both the eigenvalues $\mu^+(p_x, p_y)$ and $\mu^-(p_x, p_y) = \mu^-(p_x, p_y)^*$ survive and do not contribute to the complex phase of the partition function. 
For $p_x=0$, only one of $\mu^+(0, p_y)$ and $\mu^-(0, p_y)$ survives and contributes to the entire phase of the partition function as
\begin{equation}
  Z(KB,P+;m) \propto \mu^- \left(0,\frac{1}{2}\frac{\pi}{aN_{y}};m\right) \cdot \mu^+\left(0,\frac{3}{2}\frac{\pi}{aN_{y}};m\right) \cdots.\label{eq:ZKB_product}
\end{equation}
When $s=P{-}$, the roles of $\mu^+$ and $\mu^-$ are exchanged.

In contrast to the torus case, the expression Eq.~\eqref{eq:ZKB_product} involves contributions from many eigenvalues and $\beta^{\text{latt}}$ is not integer-valued at finite $a$ and $m$.
Nevertheless, after some calculation, we show that the $\mathrm{U}(1)$ part of the right hand side approaches to a quantized phase
\begin{equation}
  \begin{cases}
  +1 & (m>0) \\ -i & (m<0),
 \end{cases}
\end{equation}
in the continuum and large mass limit.
This gives the lattice ABK invariant
\begin{align}
  \beta^{\mathrm{latt}}(KB,P\pm;m) &= \mp2, \\
  \beta^{\mathrm{latt}}(KB,A\pm;m) &= 0,   
\end{align}
which agrees with the $\beta$ in continuum theory.
Indeed, evaluating $\beta^\mathrm{latt}$ for several $N=N_x=N_y$, we obtain values that is very close to integer, as listed in Tab.\ref{tab:KB}.
\begin{table}[htbp]
\centering

\label{tab:KB}
\begin{tabular}{@{}c cc@{}} 
\toprule
$N_x\times N_y$ & $\beta^{\mathrm{latt}}(KB,P+;m)$ & error $|\beta^{\mathrm{latt}}-\beta|$  \\ 
\midrule
$10 \times 10$ & $-1.99751\dots$ & $2.49 \times 10^{-3}$  \\ 
$20 \times 20$ & $-1.99999757\dots$ & $2.43 \times 10^{-6}$  \\ 
$30 \times 30$ & $-1.99999999762\dots$ & $2.37 \times 10^{-9}$ \\ 
\bottomrule
\end{tabular}
\caption{Numerical evaluation of $\beta^{\mathrm{latt}}(KB,P+)$ at 
$|ma|=1$ and its error from the continuum value $\beta=-2$.}
\label{tab:KB}
\end{table}
\subsection{Numerical computation}
\begin{figure}[htbp]
  \centering
  \includegraphics[width=0.7\linewidth]{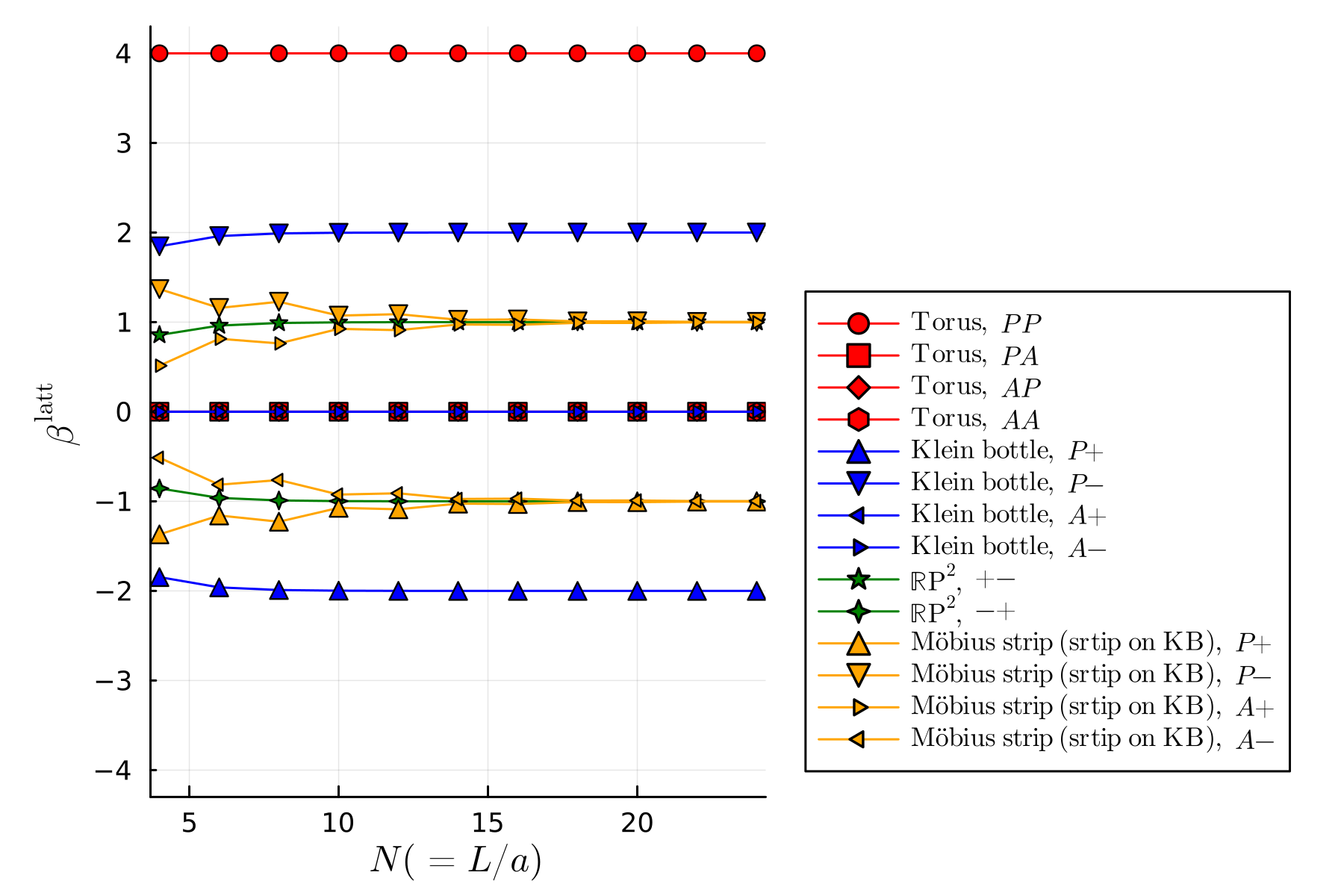} 
  \caption{The lattice ABK invariant $\beta^\mathrm{latt}$ is plotted as 
a function of $N=L/a$ for various manifolds.}
  \label{fig:batas_num}
\end{figure}

In this subsection, we numerically evaluate the ABK invariant for more general setups including the case where the momentum decomposition is not valid.
We therefore compute $\beta^\text{latt}$ by numerically evaluating the Pfaffian.
Fig.~\ref{fig:batas_num} shows the computed $\beta^\text{latt}$ for various manifold topologies and their associated $\pinm$ structures.
We set mass $m=-1/a$ and evaluate the system for varying lattice sizes $N=N_x=N_y$. At fixed manifold size $L=Na$, increasing $N$ corresponds to taking the continuum limit and the large-mass limit simultaneously.

It is obvious from the plot that the values of
$\beta^\mathrm{latt}$ are quantized
into discrete integer values, which becomes stable already at $N\sim 10$.
All the obtained results at large $N$ are consistent with the known values in the continuum theory.
It is remarkable that the data for $\RPt$ show correct values even though it is not a flat manifold and there are points where the usual square-lattice network breaks down (see the corners of Fig.~\ref{fig:manifolds_identification} (c) ).
We also verify that our domain-wall mass term construction successfully describes the topology of Möbius strip as an open manifold; there are two $\pinm$ structures and they have $\beta=\pm1$.

\section{Summary}
In this article, we have proposed a lattice formulation of the ABK invariant $\beta^\mathrm{latt}$ using the Pfaffian of the Wilson Dirac operator.
We extend Wilson fermions to manifolds with nontrivial topology including non-orientable cases.
Using orientation-reversing boundary conditions, we realize the torus, Klein Bottle and real projective plane on the lattice with their relevant $\pinm$ structures.
In addition, we have realized the Möbius strip as an open manifold via a domain-wall mass term.
By analysis based on eigenvalues (for the torus and the Klein bottle) and by numerical calculation (for the other settings), we verify that our lattice ABK invariant is quantized to high precision and correctly reproduces the continuum theory results.

 Here are additional comments for the outlook about possible
applications of this work.
We expect that our approach can be extended to other, similar topological invariants and their associated fermion theories. For example, a $\ZZ_{16}$-valued 
invariant that appears in the four dimensional Majorana fermion theory with $\pinp$ structures should be captured by this approach although the numerical cost may be high.
In this work, we have focused on free Majorana fermion systems.
It would be interesting to investigate what role $\beta^\mathrm{latt}$ plays
in the interacting theories, in particular, when the number of fermions is eight. In this case, trivialization of the SPT phase occurs, and it may be possible to gap out the fermions on the domain wall by interactions. Our lattice formulation would offer a nonperturbative tool to explore this issue.
\acknowledgments

We thank Mikio Furuta, Naoto Kan, Shinichiroh Matsuo and Kazuya Yonekura for useful discussion.
This work was
supported in part by JSPS KAKENHI Grants No.
JP21K03574, JP23K03387, JP23K22490 and JP25K07283, Japan.
This work was also supported in part by JST SPRING Grants No.
JPMJPS2138, Japan.

\small{\bibliographystyle{utphys}%
\bibliography{main_bibs.bib}}

\end{document}